\documentstyle[epsbox,12pt]{article} 
 
\begin{document}
 
\newcommand{\beq}{\begin{equation}}
\newcommand{\eeq}{\end{equation}}
\newcommand{\barr}{\begin{eqnarray}}
\newcommand{\earr}{\end{eqnarray}}

\newcommand{\andy}[1]{ }
 
\def\a{\alpha_0} \def\da{\delta\alpha}
\def\Da{D_\alpha}
\def\h{\widehat}
\def\t{\widetilde}
\def\cH{{\cal H}}
\def\bmom{\mbox{\boldmath $\omega$}}
\def\bmR{\mbox{\boldmath $R$}}

 
\def\ask{\marginpar{?? ask:  \hfill}}
\def\fin{\marginpar{fill in ... \hfill}}
\def\note{\marginpar{note \hfill}}
\def\check{\marginpar{check \hfill}}
\def\discuss{\marginpar{discuss \hfill}}
 
 
\author{ Hiromichi NAKAZATO,$^{(1)}$  
 Mikio NAMIKI,$^{(1)}$  \\ 
Saverio PASCAZIO$^{(2)}$ and
 Helmut RAUCH$^{(3)}$  \\
   \\
        $^{(1)}$Department of Physics, Waseda University \\
Tokyo 169, Japan \\
        $^{(2)}$Dipartimento di Fisica,
Universit\^^ {a} di Bari \\
and Istituto Nazionale di Fisica Nucleare, Sezione di Bari \\
I-70126  Bari, Italy  \\
        $^{(3)}$Atominstitut der \"Osterreichischen
Universit\"aten  \\
A-1020 Wien, Austria}
 
\title{ Understanding the quantum Zeno effect }
 
\date{}
 
\maketitle
 
\begin{abstract}
The quantum Zeno effect consists in the hindrance of the evolution
of a quantum system that is very frequently monitored and {\em found} 
to be in its initial state at {\em every} single measurement. 
On the basis of the correct formula for the survival probability, i.e.\ the
probability of finding the system in its initial state at {\em every} single
measurement, we critically analyze a recent proposal and experimental test, 
that make use of an oscillating system.
\end{abstract}

\vspace*{.5cm} PACS: 03.65.Bz, 42.50.-p \vspace*{.5cm}

\newpage
 

The seminal formulation of the quantum Zeno effect, due to Misra and Sudarshan
\cite{Misra}, deals with the probability of observing an unstable system
in its initial state {\em throughout} a time interval $\Delta = [0,t]$.
The purpose of this note is to point out that the quantum Zeno effect 
has not been experimentally observed, yet, in its original formulation. 
Indeed, we shall argue that the interesting 
proposal by Cook \cite{Cook}, that makes use of a two-level system
undergoing Rabi oscillations, as well as the beautiful experiment performed 
by Itano et al.\ \cite{Itano1}, investigate the probability 
of finding the initial state {\em at} time $t$, {\em regardless}
of the actual state of the system in the time interval $\Delta$.
As we shall see, in general, if the temporal behavior of the system is 
oscillatory, this  probability includes the possibility that 
transitions of the type: initial state $\rightarrow$ other states 
$\rightarrow$ initial state, actually take place.
Of course, this remark does not invalidate the soundedness of the analysis
in \cite{Cook} and of the experiment \cite{Itano1}. 

The temporal behavior of quantum mechanical systems is a long-standing issue
of investigation \cite{Hell} (for a review and a collection of 
recent developments, see \cite{IJMPB}),
and the curious features of the short-time behavior 
of the so-called ``survival" probability of a quantum mechanical state,
leading to what was to be named ``quantum Zeno paradox" \cite{Misra},
were already known about 30 years ago \cite{stbeh}.
However, renewed interest in the above topic was motivated by 
Cook's idea \cite{Cook} and its subsequent experimental verification
\cite{Itano1}. The experiment by Itano et al.\ 
provoked a lively debate \cite{Petrosky,qze,Mensky},
that has essentially focussed on two aspects of the problem.
{}First, it has been shown, and it is now becoming a widespread viewpoint,
that the experimental results can be explained
by making use of a unitary dynamics \cite{qzePN,qze}.
Notice that an analogous point was raised by Peres quite a few 
years ago \cite{Peres}. 
Second, it has been argued that the so-called limit of continuous observation
is in contradiction with Heisenberg's uncertainty principle
and does not take into account unavoidable quantum mechanical losses,
and is therefore to be considered unphysical \cite{NNPR}.

Nowadays most physicists tend to view this phenomenon as a purely 
dynamical process, 
in which von Neumann's projections can be substituted by
spectral decompositions \cite{Wigner,np}, so that the phase correlation 
among different branch waves is perfectly kept.
{}For this reason, one often speaks of quantum Zeno {\em effect}
(QZE) \cite{Petrosky,qze}, rather than quantum Zeno {\em paradox} 
\cite{Misra}.

However, surprisingly, nobody seems to have realized that,
strictly speaking, Cook's proposal and Itano et al.'s experiment
are conceptually at variance with the original formulation of the QZE.
Misra and Sudarshan, in their seminal paper \cite{Misra}, 
endeavoured to define ``the probability ${\cal P}(0,T;\rho_0)$ that no decay
is found {\em throughout the interval} $\Delta = [0,T]$ when the initial 
state of the system was known to be $\rho_0$."
(Italics in the original. Some symbols have been changed.)
The definition given in Ref.\ \cite{Misra} is 
\andy{Pdef}
\beq
{\cal P}(0,T;\rho_0) \equiv \lim_{N \rightarrow \infty} P^{(N)}(0,T;\rho_0) ,
\label{eq:Pdef}
\eeq
where $P^{(N)}(0,T;\rho_0)$ is the probability of observing 
the initial state $\rho_0$ in a series of $N$ observations,
performed at times $t_n = nT/N, \; (n=1, \ldots N)$,
in order to ascertain whether the system is still undecayed.

In order to facilitate comprehension of the following analysis, 
it is worth stressing that the above-mentioned ``survival probability" 
of the initial state $\rho_0$ is the 
probability of finding the system under investigation
in $\rho_0$ at {\em every} measurement, 
during the interval $\Delta$. This is a subtle point, as we shall see.

{}For the sake of clarity, we shall first carefully analyze Itano et al.'s 
derivation of what they interpreted as a realization of the QZE, and then 
scrutinize Cook's formulae.
Consider a three-level atomic system, on which an rf field of frequency
$\omega$ provokes Rabi oscillations between levels 1 and 2. 
In the rotating wave approximation and in absence of detuning,
the equations of motion for the density matrix $\rho_{ij}\; (i,j=1,2)$ 
read 
\andy{Rabieq}
\barr
\dot \rho_{11} & = & i \frac{\omega}{2} (\rho_{21} - \rho_{12}),
     \nonumber \\
\dot \rho_{12} & = & i \frac{\omega}{2} (\rho_{22} - \rho_{11}),
                  \label{eq:Rabieq} \\
\dot \rho_{22} & = & i \frac{\omega}{2} (\rho_{12} - \rho_{21}),
     \nonumber 
\earr
where the dot denotes derivative with respect to time.

By applying a technique invented by Feynman, Vernon and Hellwarth 
\cite{FVH}, one can recast the above equations of motion in a very simple 
form, in which the use of rotating coordinates, introduced by 
Block \cite{Block} and Rabi, Ramsey and Schwinger \cite{RRS}, turns out
to be particularly advantageous.
Define 
\andy{FWHdef}
\barr
R_1 & \equiv & \rho_{21} + \rho_{12},
     \nonumber \\
R_2 & \equiv & i (\rho_{12} - \rho_{21}),
                  \label{eq:FWHdef} \\
R_3 & \equiv & \rho_{22} - \rho_{11} \equiv P_2-P_1,
     \nonumber 
\earr
where $P_j \equiv \rho_{jj}$ is the probability that the atom
is in level $j \; (j=1,2)$. 
Since $P_1+P_2=1$, one gets 
\andy{PR}
\beq
P_2=\frac{1}{2}(1+R_3). 
\label{eq:PR}
\eeq
In terms of the quantities $\bmR \equiv (R_1,R_2,R_3)$ and 
$\bmom \equiv (\omega,0,0)$, Eqs.\ (\ref{eq:Rabieq}) become
\andy{Rform}
\beq
\dot{\bmR} = \bmom \times \bmR .
\label{eq:Rform}
\eeq
The solution of the above equation, 
with initial condition $\bmR (0) \equiv (0,0,-1)$ (only level 1 is 
initially populated) reads
\andy{Rsol}
\beq
\bmR (t) = (0, \sin \omega t, -\cos \omega t).
\label{eq:Rsol}
\eeq
If the transition beween the two levels is driven by an 
on-resonant $\pi$ pulse, of duration $T=\pi/\omega$,
one gets $\bmR (T) \equiv (0,0,1)$, so that $\rho_{22} =1,
\rho_{11}=0$, and only level 2 is populated at time $T$.

The reasoning of Ref.\ \cite{Itano1} is the following.
Assume you perform a measurement at time $\tau=\pi/N \omega =T/N$, by shining
on the system a very short ``measurement" pulse, that provokes 
transitions from level 1 to level 3, with subsequent spontaneous emission 
of a photon.\footnote{We are not addressing the (delicate) point that a
measurement pulse, however short, must have a certain {\em finite} time 
duration. As a consequence, one must take into account the inevitable
spread $\triangle \omega$ of the measurement pulse, and modify accordingly the 
following formulae. This problem is a very subtle one and will be properly
addressed in a forthcoming paper \cite{NNPR2}.} 
The measurement pulse ``projects" the atom into level 1 or 2
(``naive wave function collapse"). Because a
measurement ``kills" the off-diagonal terms $\rho_{12}$ and $\rho_{21}$
of the density matrix, while leaving unaltered its diagonal terms
$\rho_{11}$ and $\rho_{22}$, one obtains
\andy{Rsolt}
\beq
\bmR (\pi/N\omega) = [0, \sin (\pi/N),-\cos (\pi/N)]
 \stackrel{{\rm measurement}}{\longrightarrow} [0,0,-\cos (\pi/N)]
 \equiv \bmR^{(1)}.
\label{eq:Rsolt}
\eeq
Then the evolution restarts, according to Eq.\ (\ref{eq:Rform}),
but with the new initial condition $\bmR^{(1)}$. After $N$ measurements,
at time $T=N\tau=\pi/\omega$,
\andy{RT}
\beq
\bmR (T) = [0,0,-\cos^N (\pi/N)]
 \equiv \bmR^{(N)}.
\label{eq:RT}
\eeq
The probabilities that the atom is in level 2 or 1 at time $T$,
after the $N$ measurements, are therefore given by [see
Eq.\ (\ref{eq:PR})]
\andy{P2,P1}
\barr
P_2^{(N)}(T) & = & \frac{1}{2} \left[ 1+R_3^{(N)} \right] = 
\frac{1}{2} \left[1 - \cos^N (\pi/N) \right],
\label{eq:P2} \\
P_1^{(N)}(T) & = & 1 - P_2^{(N)}(T) =
\frac{1}{2} \left[1 + \cos^N (\pi/N) \right] ,
\label{eq:P1} 
\earr
respectively.
Since $P_2^{(N)}(T) \rightarrow 0$ and $P_1^{(N)}(T) \rightarrow 1$ 
as $N \rightarrow \infty$,
this is interpreted as quantum Zeno effect.\footnote{The $N\rightarrow
\infty$ limit
is in contradiction with Heisenberg uncertainty principle, and is 
therefore unphysical. It is possible to set a physical limit on the maximum 
value that $N$ can attain in a certain experimental situation 
\cite{NNPR,NNPR2}.}
The experimental result are in very good agreement with the above formulae.
However, this is {\em not} the quantum Zeno effect {\em \`a
la} Misra and Sudarshan: Equation
(\ref{eq:P2}) [(\ref{eq:P1})] expresses only the probability that the atom
is in level 2 [1] at time $T$, after $N$ measurements,
{\em independently} of its past history. In particular, Eqs.\
(\ref{eq:P2})-(\ref{eq:P1}) take into account the 
{\em possibility that one level gets repopulated after the atom
has made transitions to the other level}. In order to shed light on this
very important (and rather subtle) point, let us look explicitly
at the first two measurements.

After the first measurement, $\bmR^{(1)}$ is given by Eq.\ (\ref{eq:Rsolt}) 
and 
\andy{R31}
\beq
R_3^{(1)} = - \cos \frac{\pi}{N} = P_2^{(1)} - P_1^{(1)},
\label{eq:R31}
\eeq
where $P_j^{(1)}$ is the occupation probability of level $j \; (j=1,2)$
at time $\tau=\pi/N\omega$, after the first measurement pulse:
\andy{probb2,1}
\barr
P_2^{(1)} & = & 
\frac{1}{2} \left( 1 + R_3^{(1)} \right) 
 = \sin^2 \frac{\pi}{2N} ,   \label{eq:probb2} \\
P_1^{(1)} & = & 1 - P_2^{(2)} = \cos^2 \frac{\pi}{2N} .
  \label{eq:probb1}
\earr
After the second measurement, one obtains
\andy{R32}
\beq
R_3^{(2)} = - \cos^2 \frac{\pi}{N} = P_2^{(2)} - P_1^{(2)},
\label{eq:R32}
\eeq
where the occupation probabilities at time $2\tau= 2 \pi/N \omega$ read
\andy{pp2,1}
\barr
P_2^{(2)} & = & 
\frac{1}{2} \left( 1 + R_3^{(2)} \right)  
 = 2 \sin^2 \frac{\pi}{2N} \cos^2 \frac{\pi}{2N} ,
  \label{eq:pp2} \\
P_1^{(2)} & = & 1 - P_2^{(2)} =
    \cos^4 \frac{\pi}{2N} + \sin^4 \frac{\pi}{2N} .
  \label{eq:pp1}
\earr
It is then obvious that $P_1^{(2)}$, in Eq.\ (\ref{eq:pp1}), 
is {\em not the survival probability} of level 1,
according to the seminal definition (\ref{eq:Pdef}).
It is just the probability that level 1 is populated at time $t=2\pi/N \omega$,
including the possibility that the 
transition $1 \rightarrow 2 \rightarrow 1$ took place, with probability
$\sin^2 \frac{\pi}{2N} \cdot \sin^2 \frac{\pi}{2N} = \sin^4 \frac{\pi}{2N}$. 
By contrast, the {\em survival} probability, namely the probability
that the atom is found in level 1 {\em both} in the first and second
measurements, is given by $P_1^{(1,2)}= \cos^2 \frac{\pi}{2N} \cdot
\cos^2 \frac{\pi}{2N} = \cos^4 \frac{\pi}{2N}$.
{}Figure~1 shows what happens during the first two measurements in the 
experiment \cite{Itano1}. 
\begin{figure}
\vspace*{12pt}
\begin{center}
\epsfile{file=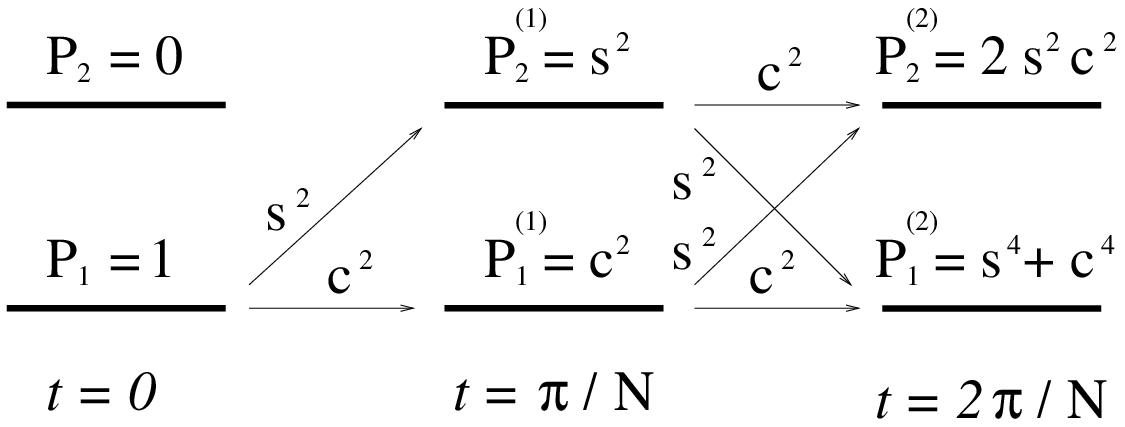}
\caption{Transition probabilities after the first two measurements
($s=\sin \frac{\pi}{2N}$ and $c=\cos \frac{\pi}{2N}$).}
\end{center}
\end{figure}

In the general case, after $N$ measurements, the probability that level 1 is 
populated at time $T$, independently of its ``history", is given by 
(\ref{eq:P1}), and includes the possibility that transitions to level 
2 took place. As a matter of fact, it is not difficult to realize that 
(\ref{eq:P2})-(\ref{eq:P1}) conceal a binomial distribution:
\andy{bindis}
\barr
\sum_{n \; {\rm even}} 
\left(    \begin{array}{c} N \\ n  \end{array}
\right)
s^{2n}c^{2(N-n)} & = & c^{2N} 
\sum_{n \; {\rm even}} 
\left(    \begin{array}{c} N \\ n  \end{array}
\right) 
(s/c)^{2n}  \nonumber \\
 & = &
\frac{c^{2N}}{2} \left[
\sum_{n=0}^N  \left( \begin{array}{c} N \\ n  \end{array} \right) (s/c)^{2n} + 
\sum_{n=0}^N  \left( \begin{array}{c} N \\ n  \end{array}
\right) (-1)^n(s/c)^{2n}
\right]   \nonumber \\
 & = & 
\frac{c^{2N}}{2} \left[ (1+(s/c)^2)^N + (1-(s/c)^2)^N \right] \nonumber \\
 & = & \frac{1}{2} \left[1 + \cos^N (\pi/N) \right] \nonumber \\
 & = & P_1^{(N)}(T) = 1 - P_2^{(N)}(T)  ,
  \label{eq:bindis}
\earr
where $\sum_{n \; {\rm even}}$ is a sum aver all even values of $n$ between 0
and $N$, $s=\sin (\pi/2N), c=\cos (\pi/2N)$.\footnote{Mensky
\cite{Mensky} first noticed the 
occurrence of a binomial distribution in connection with the QZE for 
oscillating system, without however pointing out the discrepancy with Misra
and Sudarshan's definition of survival probability. The result 
(\ref{eq:bindis}) is, to our knowledge, new.}
Therefore 
\andy{Pdiscl2,1}
\barr
P_2^{(N)}(T)  & = & 1 - \sum_{n \; {\rm even}} 
\left(    \begin{array}{c} N \\ n  \end{array}
\right) \sin^{2n} \frac{\pi}{2N} \cos^{2(N-n)} \frac{\pi}{2N} ,
  \label{eq:Pdiscl2}    \\
P_1^{(N)}(T)  & = & \sum_{n \; {\rm even}} 
\left(    \begin{array}{c} N \\ n  \end{array}
\right) \sin^{2n}\frac{\pi}{2N} \cos^{2(N-n)} \frac{\pi}{2N} ,
  \label{eq:Pdiscl1}   
\earr
which clearly shows that Eqs.\ (\ref{eq:P2})-(\ref{eq:P1})  or 
(\ref{eq:Pdiscl2})-(\ref{eq:Pdiscl1}) include all possible
transitions between levels 1 and 2, in such a way that at time $T$ the system 
is, say, in level 1 after having made an even number 
($n=0, 2,\ldots$, etc.) of transitions between levels 1 and 
2. It should be clear now that the result (\ref{eq:P1})
is conceptually very different from Misra and Sudarshan's
survival probability (\ref{eq:Pdef}). The correct formula for the survival
probability, in the present case, 
is obtained by considering {\em only} the $n=0$ term in  (\ref{eq:Pdiscl1}):
\andy{pcorr1}
\beq
{\cal P}_1^{(N)}(T) = \cos^{2N} \frac{\pi}{2N} .
  \label{eq:pcorr1} 
\eeq
Equation (\ref{eq:pcorr1}) is just the ``survival probability", namely the
probability that level 1 is populated at every 
measurement, at times $n\tau=nT/N$ ($n=1,\ldots,N$).\footnote{
Equation (\ref{eq:pcorr1}) was first given in Section V of Ref.\ \cite{qzePN}
(see in particular footnote 21).} 

A comparison with the formulae of Ref.\ \cite{Itano1} is not straightforward,
due to the fact that the authors analyzed their results in terms 
of the quantity $P_2^{(N)}(T)$, rather than $P_1^{(N)}(T)$.
At any rate, Eq.\ (\ref{eq:pcorr1}) implies
\andy{pcorr2}
\beq
{\cal P}_2^{(N)}(T) = 1 - \cos^{2N} \frac{\pi}{2N} .
  \label{eq:pcorr2}  
\eeq
Equation (\ref{eq:pcorr2}) can be compared to (\ref{eq:P2}):
Even though they tend to the same limiting
value 0 [in either case $\sin (\pi/2N) \rightarrow 0$ as $N
\rightarrow \infty$], they give different results, in particular when
$N$ is small, as shown in Table 1.
\begin{center}
{\bf Table 1} \\  \quad \\
\begin{tabular}{|l|c|c|c|c|c|c|c|}      \hline
$ N $        &  1$^\dagger$ &  2  &  4  &  8  &  16  &  32 &  64 \\ 
  \hline\hline
$P_2^{(N)}(T)$ 
              & 1  &  .5 &  .3750 & .2346 & .1334 & .0716 & .0371 \\
 \hline
${\cal P}_2^{(N)}(T) $
              & 1 & .75 & .4692 & .2668 & .1431 & .0742 & .0378 \\
  \hline
\end{tabular} \\ \quad  \\
$^\dagger$ $N=1$ means that only a final measurement is performed, 
at time $T$.
\end{center}

It must be emphasized that we are not criticing the soundedness of the
nice experiment 
\cite{Itano1}. Indeed, the experimental results obtained by
Itano et al.\ are in excellent agreement with Eqs.\ (\ref{eq:P2}) or
(\ref{eq:Pdiscl2}).
We only claim that this experiment, although correctly performed,
is conceptually at variance with the original idea
on the QZE, as defined by Misra and Sudarshan, because the right
expression for the survival probability,
according to (\ref{eq:Pdef}), is given by (\ref{eq:pcorr1})
and not by (\ref{eq:Pdiscl1}).

Let us now look at Cook's derivation of the QZE. For the sake of clarity,
we shall present his analysis in a slightly simplified case.
Starting from the set of equations (\ref{eq:Rabieq}), Cook obtained the 
following rate equations
\andy{Cook1,2}
\barr
\dot P_1 & = & k (P_2 -P_1)
  \label{eq:Cook1} \\
\dot P_2 & = & k (P_1 -P_2)
  \label{eq:Cook2}
\earr
where $k = \omega^2 \tau/2$, $\tau$ being the time interval between 
measurement pulses. These equations yield, at time $T=\pi/\omega$
\andy{Cookres}
\beq
P_2(T) =  \frac{1}{2} \left[1 - \exp (-\pi^2/2N) \right].
  \label{eq:Cookres}
\eeq
(A misprint in Ref.\ \cite{Cook} has been corrected.)
The above formula is interpreted as a quantum Zeno effect.
Once again, this is not correct in a strict sense: The above equation
expresses the occupation probability of level 2, independently
of its history. Clearly, the rate equations 
(\ref{eq:Cook1})-(\ref{eq:Cook2}) take into account the possibility
of transitions $1\rightarrow 2 \rightarrow 1$, and so on, and therefore
cannot be viewed as expressing ``survival" probabilities,
as in Eq.\ (\ref{eq:Pdef}).
It should be stressed that the conclusions drawn in this Letter hold true for 
all those 
situations in which the temporal behavior of the system under investigation 
is of the oscillatory type, and no precautions are taken in order to prevent
repopulation of the initial state.

{}Finally, it is worth briefly commenting on the $N\rightarrow \infty$
limit (continuous observation). It was shown \cite{NNPR,NNPR2} that 
this limit is unphysical, for it is in contradiction with
Heisenberg's uncertainty principle, and set a reasonable physical
limit for the maximum value that $N$ can attain in an experimental
test of the QZE involving neutron spin. Venugopalan and Ghosh
\cite{Anu} criticized this result on the basis of an analysis whose
starting point was Eq.\ (\ref{eq:Cookres}). However, as we have seen, 
(\ref{eq:Cookres}) is {\em not} related to the survival probability,
according to the definition (\ref{eq:Pdef}), so that the calculation 
of Ref.\ \cite{Anu}, although mathematically correct, is not physically
relevant for our problem. Incidentally, in the light of our analysis, it is not
surprising that the authors of Ref.\ \cite{Anu}, 
by applying the uncertainty principle, obtained the limiting
value $P_2(T) \rightarrow 1/2$, in the large-$N$ limit, from
Eq.\ (\ref{eq:Cookres}). Such a result is to be expected,
on the basis of Cook's equations (\ref{eq:Cook1})-(\ref{eq:Cook2}),
but refers to a physically different situation, not to the QZE.

In conclusion, we would like to put forward a few remarks.
The real problem related with Cook's proposal and Itano et al.'s
experiment is that the state of the atom is {\em not} observed
at intermediate times. As a matter of fact, its observation
would probably rise difficult technical problems, for one
should be able to ``select", after each measurement pulse,
{\em which} atoms are in level 1 and discard those atoms 
that are in level 2.

The quantum theory of measurement \cite{von,Zurek}
is still full of pitfalls and conceptual 
difficulties. One has to be extremely careful when applying 
von Neumann's projection postulate. A quantum measurement implies 
the occurrence of decoherence, but the {\em vice versa}
is not necessarily true, as we have seen: It may happen
that the system is practically incoherent, but one still does not
know, in practice, {\em which} state the atom is in.

Very promising candidates for an experimental observation
of a genuine QZE seem to be those experiments involving neutron spin
\cite{NNPR} or photon polarization \cite{Zeilinger}.
There is certainly more to come, on this fascinating subject.

\vspace*{.7cm}
We thank C.\ Presilla for bringing Ref.\ \cite{Mensky} to our attention.
M.N.\ was partially supported by the Japanese Ministry of Education,
Science and Culture,
and S.P.\ by the Japanese Society for the Promotion of Science, under a
bilateral exchange program with Italian Consiglio Nazionale 
delle Ricerche, and by the Administration Council of the University of Bari.
S.P. thanks the High Energy Physics Group of Waseda University
for their kind hospitality and H.N. acknowledges the kind hospitality 
at the Department of Physics, University of Bari.


 
 
\end{document}